\documentclass[a4paper,fleqn]{cas-sc}
\usepackage{diagbox}
\usepackage[authoryear]{natbib}
\usepackage{wrapfig}
\usepackage[ruled,vlined]{algorithm2e}
\usepackage{algpseudocode} 
\usepackage{graphicx}
\graphicspath{ {Figures/} }
\usepackage{caption}
\usepackage{subcaption}
\usepackage{longtable}

\def\tsc#1{\csdef{#1}{\textsc{\lowercase{#1}}\xspace}}
\tsc{WGM}
\tsc{QE}
\tsc{EP}
\tsc{PMS}
\tsc{BEC}
\tsc{DE}

\begin{document}
\let\WriteBookmarks\relax
\def\floatpagepagefraction{1}
\def\textpagefraction{.001}
\shorttitle{}

\title [mode = title]{Self-Citations in Academic Excellence: Analysis of the Top 1\% Highly Cited India-Affiliated Research Papers}                      
\author[1,2]{Kiran Sharma}\corref{cor1}
\ead{kiran.sharma@bmu.edu.in}
\author[3]{Parul Khurana}
\address{School of Engineering \& Technology, BML Munjal University, Gurugram, Haryana-122413, India }
\address{Center for Advanced Data and Computational Science, BML Munjal University, Gurugram, Haryana-122413, India }
\cortext[cor1]{Corresponding author}
\address{School of Computer Applications, Lovely Professional University, Phagwara, Punjab-144411, India }
\begin{abstract}
Citations demonstrate the credibility, impact, and connection of a paper with the academic community. Self-citations support research continuity, but, if excessive, may inflate metrics and raise bias concerns. The aim of the study is to examine the role of self-citations towards the research impact of India. To study this,  3.58 million papers affiliated with India from 1947 to 2024 in the Scopus database were downloaded, and 2.96 million were filtered according to document type and publication year up to 2023. Further filtering based on high citation counts identified the top 1\% of highly cited papers, totaling 29,556. The results indicate that the impact of Indian research, measured by highly cited papers, has grown exponentially since 2000, reaching a peak during the 2011–2020 decade. Among the citations received by these 29,556 papers, 6\% are self-citations. Papers with a high proportion of self-citations (>90\%) are predominantly from recent decades and are associated with smaller team sizes. The findings also reveal that smaller teams are primarily domestic, whereas larger teams are more likely to involve international collaborations. Domestic collaborations dominate smaller team sizes in terms of both self-citations and publications, whereas international collaborations gain prominence as team sizes increase. The results indicate that while domestic collaborations produce a higher number of highly cited papers, international collaborations are more likely to generate self-citations. The top international collaborators in highly cited papers are the USA, followed by UK, and Germany.
\end{abstract}


\maketitle
\section{Introduction}

Citations are essential in academic research, acknowledging previous work, demonstrating integrity, and situating new studies within a broader scientific context. They serve as a key metric for assessing the impact of research, with high citation counts reflecting significant contributions to the field. Furthermore, citations facilitate knowledge dissemination, foster collaboration, and link studies between disciplines~\citep{bornmann2008citation}. However, self-citations, where authors cite their own work, provide continuity by linking new findings to prior contributions, especially in cumulative research. They also increase the visibility of newly published papers, which can attract external citations by highlighting related work~\citep{hyland2003self}. However, excessive self-citations can artificially inflate the citation metrics, misrepresenting the true influence of the paper, and raising concerns about bias~\citep{fowler2007does}. Thus, while citations and self-citations are vital tools for measuring academic impact, their appropriate use is essential to maintain credibility and transparency in research.

Moreover, when self-citations are used in an excessive or strategic manner to inflate citation metrics, it distorts the author's as well as the organization's academic influence~\citep{moed2006citation}. Today, citations in the form of scientific influence are used by various academic and government organizations for hiring, promotions, institutional prestige, bridging knowledge across different fields, fostering interdisciplinary research and funding decisions~\citep{van2013bibliometric}. In such cases, self citations may create citation loops to potentially skew critical bibliometric indicators~\citep{tahamtan2019citation}. The existence of groups in the form of ``citation cartels'' also engage in reciprocal citation practices, which further compounds the issue~\citep{hillman2019quality}. This trend underscores the urgent need for a nuanced understanding of self-citations across different academic backgrounds as high citations indicate that the particular study has substantial contribution in the field of research~\citep{hirsch2005index}.

As the scientific community understood the elevation, narrative, and opportunistic power of self-citations, concerns arose about the ethical implications of artificial citations~\citep{van2019hundreds}. Some argued in favor of self-citations, stating it as a reflection of specialization, while others presented them as manipulations ~\citep{costas2010self}. The tipping point came when researchers unravel the self-citation patterns at the level of authors, country exhibitions, and academic organizations~\citep{hellsten2007self}.

Citations in the academic world work as the thread that weaves the vast fabric of human knowledge. If utilized properly and for the advancement of the community, they are more than just numbers~\citep{hodge2025assessing, szomszor2020much}. They enhance human knowledge by guaranteeing coherence and continuity. Institutions and financial agencies need to take into account the caliber of contributions rather than just the quantity of citations~\citep{hussein2024self}. Most significantly, the scientific community needs to keep improving its evaluation methods so that a researcher's effect is determined by academic merit rather than metric manipulation~\citep{martin2013whither}.

\section{Research objectives}
The study examines the top 1\% of highly cited papers and aims to achieve the following objectives:
\begin{enumerate}
 \item Evaluate the concentration of self-citations in the highly cited papers over year and decades.
 \item Investigate the influence of team size on self-citation patterns.
\item Explore the impact of domestic and international collaborations on highly cited papers and the associated concentration of self-citations.
\end{enumerate}
\section{Methodology}
\begin{figure}[!h]
    \centering
\includegraphics[width=0.55\linewidth]{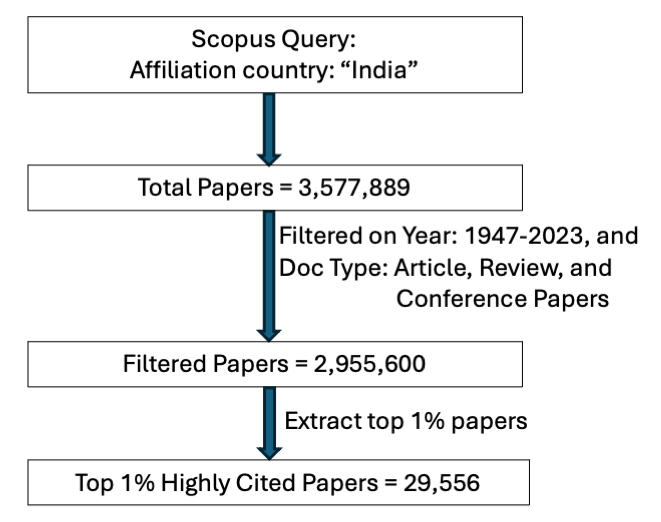} 
\caption{Search strategies undertaken to identify top 1\% highly cited papers affiliated to India.}
\label{fig:flowchart}
\end{figure}

In figure~\ref{fig:flowchart}, the flow chart outlines the process of selecting the top 1\% highly cited research papers affiliated with India, based on data retrieved from Scopus. A total of 3.58 million papers from 1947 to 2024 were downloaded from Scopus, searching for the affiliation country as ``India''. The dataset was then refined to include only articles, conference proceedings, and reviews, focusing on publications up to 2023, resulting in 2.96 million papers from 1947 to 2023. Of these, articles accounted for 2.28 million (76.29\%), conference proceedings for 0.56 million (18.94\%) and reviews for 0.14 million (4.78\%). The papers were organized in descending order based on the number of citations received. Further filtering identified the top 1\% of highly cited papers, totaling 29,556. Within this group, articles comprised 21,645 papers (73.23\%), reviews 7,100 (24\%) and conference papers 811 (2.74\%). Finally, the filtering process systematically narrowed down a massive dataset of more than 3.5 million papers to a smaller set of highly impactful publications. The top 1\% highly cited papers (29,556) represent the most influential research outputs affiliated with India, highlighting the country's global academic and scientific contributions.

\section{Results and Discussion}

\subsection{Citations vs. self-citations}
\begin{figure}[!h]
    \centering
\includegraphics[width=0.75\linewidth]{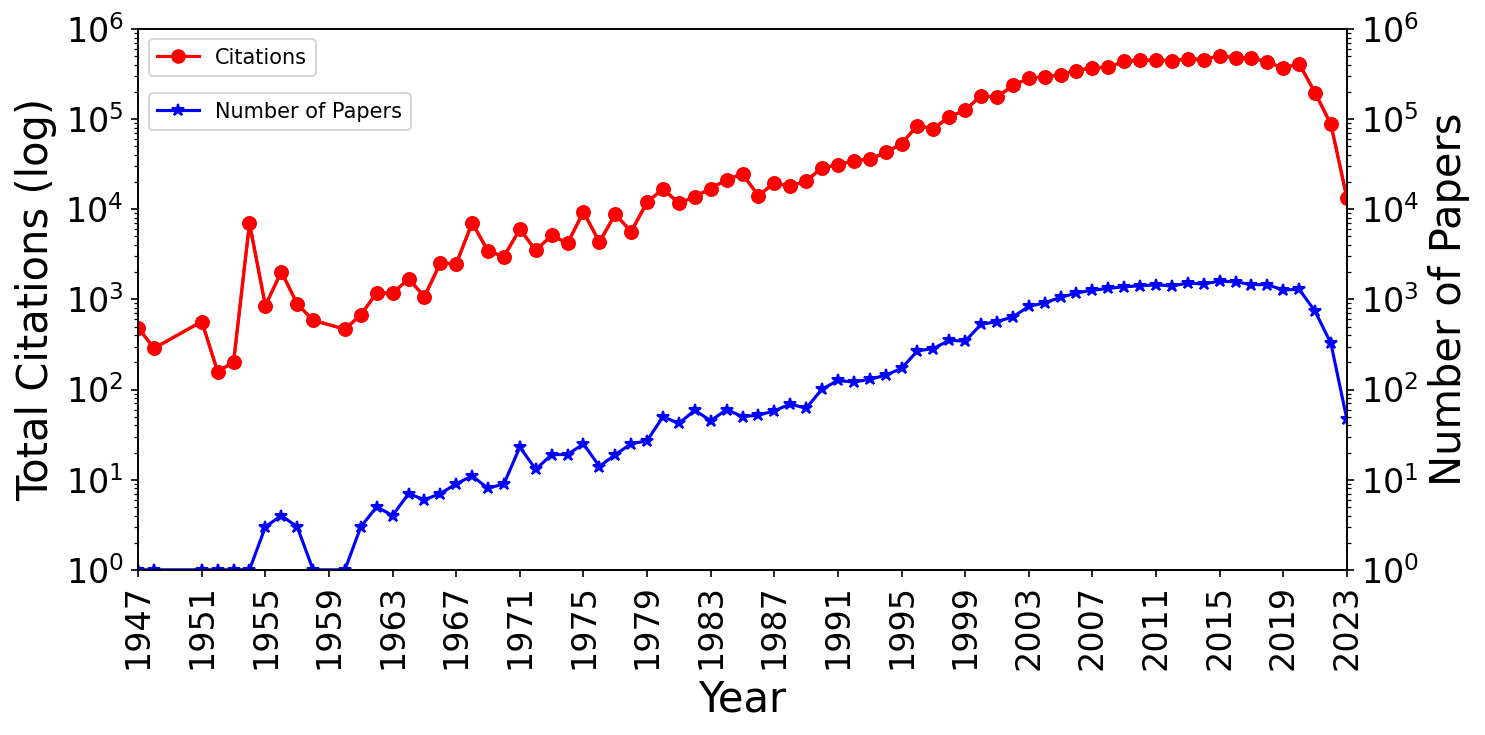} 
\caption{Yearwise trend of number of publications and total citations received.}
\label{fig:yearPub}
\end{figure}

Citations and self-citations are indispensable tools for academic research, helping to recognize prior work, measure impact, and foster scholarly communication. Striking the right balance between self-referencing and engaging with the broader academic community is essential to maintain the integrity and quality of research. Figure~\ref{fig:yearPub} represents the trends in total citations and number of papers (logarithmic scale) affiliated with India over time (from 1947 to 2023). Very few papers were published during 1947–1980 (early stage), as reflected by the flat portion of the blue star line. Citations are also minimal, but the red dotted line shows occasional spikes (possibly due to a few influential papers published during this period). The graph demonstrates India's remarkable progress in producing highly influential research papers, particularly post-2000.

\begin{figure}[!h]
    \centering
\includegraphics[width=0.75\linewidth]{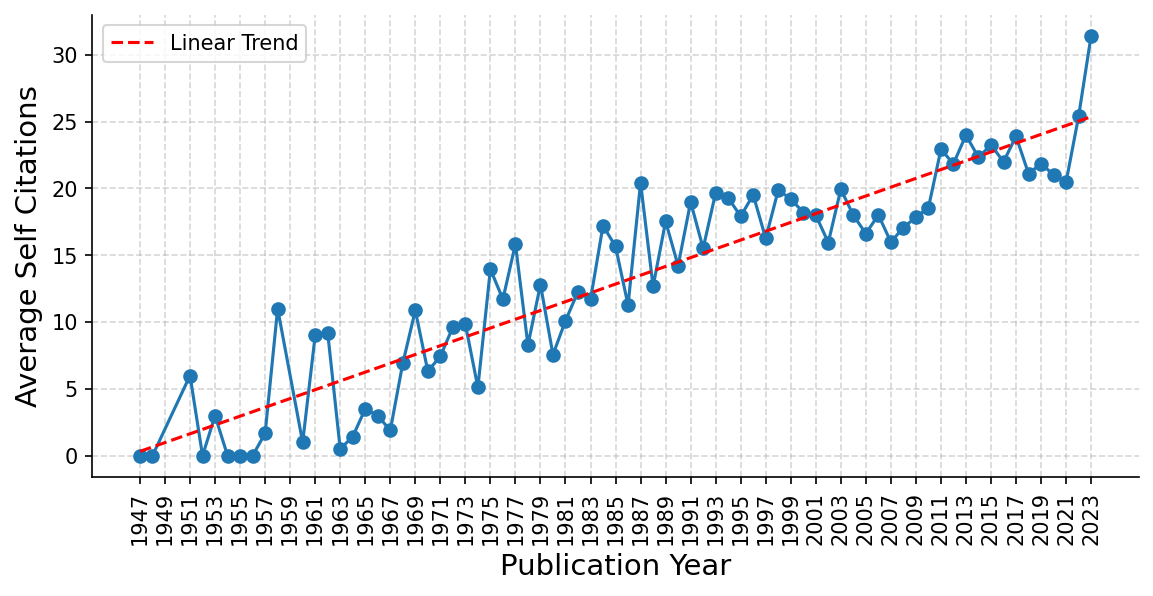} 
\caption{Average self-citations over the years. Red dashed line represents the trend line with slope 0.33.}
\label{fig:Avg_SC}
\end{figure}

A total of 29,556 highly cited papers received 97,53,620 citations in total, as shown in Table~\ref{table:totalPapers}. Of these 9.75 million citations, 0.59 million (593,321) are self-cited, that is, 6\% citations are self-citations.
Figure~\ref{fig:Avg_SC} represents the trend line of the average self-citations over years received by the highly cited papers. The data exhibit fluctuations, with periods of increase and occasional declines, but the overall trend reveals a steady growth in self-citations over time. The positive slope of the red trend line confirms this upward trajectory, suggesting that self-citations have generally become more frequent in recent years.

Table~\ref{table:totalPapers} provides an analysis of the top 1\% highly cited papers affiliated with India, organized by decades. The table includes the count of papers, their percentage contribution to the total, the total citations, share of self-citations and whether the papers were published under Open Access (Yes) or not (No). Among the 29,556 highly cited papers, 21.4\% (6,327) were published as open access, while 78.59\% (23,229) were not open access. The transition to Open Access is evident, with nearly 21\% of recent highly cited papers being openly accessible. The production of highly cited papers increased dramatically after 2000, with nearly 90\% of the top 1\% papers produced between 2001 and 2023. The most significant contribution came from the 2011–2020 decade (49.45\%). Citations reflect the growing impact of Indian research globally. Papers from 2011–2020 have received the highest citations (48,98,374), nearly half of the total. This trend demonstrates India's rising contribution to global research impact, particularly in recent decades, through both high citation counts and improved accessibility. In contrast, an analysis of the proportion of self-citations to total citations reveals that 6.5\% of self-citations refer to papers published during 2021–2023, 6.6\% to those from 2011–2020 and 5.5\% to papers from 2002–2010. Table~\ref{tab:sc} presents a list of papers in which self-citations account for more than 90\% of their total citations. These papers are primarily from recent decades and are associated with smaller team sizes.

\begin{table}[]
\caption{Decade-wise count of number of publications, total citations, self citations and paper status:open access or not.}
\label{table:totalPapers}
\begin{tabular}{|l|cc|c|c|cc|}
\hline
\multirow{2}{*}{\textbf{Decades}} & \multicolumn{2}{c|}{\textbf{TP}}                       & \multirow{2}{*}{\textbf{\begin{tabular}[c]{@{}c@{}}Total \\ Citations\end{tabular}}} & \multirow{2}{*}{\textbf{\begin{tabular}[c]{@{}c@{}}Self\\ Citations\end{tabular}}} & \multicolumn{2}{c|}{\textbf{Open Access}}       \\ \cline{2-3} \cline{6-7} 
                                  & \multicolumn{1}{c|}{\textbf{Count}} & \textbf{\%Count} &   &   & \multicolumn{1}{c|}{\textbf{Yes}} & \textbf{No} \\ \hline
1947-1950                         & \multicolumn{1}{c|}{2}              & 0.01             & 801                                                                                  & 0                                                                                  & \multicolumn{1}{c|}{0}            & 2           \\ \hline
1951-1960                         & \multicolumn{1}{c|}{15}             & 0.05             & 13,083                                                                               & 26                                                                                 & \multicolumn{1}{c|}{1}            & 14          \\ \hline
1961-1970                         & \multicolumn{1}{c|}{61}             & 0.21             & 24,585                                                                               & 364                                                                                & \multicolumn{1}{c|}{3}            & 58          \\ \hline
1971-1980                         & \multicolumn{1}{c|}{193}            & 0.65             & 76,997                                                                               & 2,325                                                                              & \multicolumn{1}{c|}{8}            & 185         \\ \hline
1981-1990                         & \multicolumn{1}{c|}{547}            & 1.85             & 1,92,674                                                                             & 8,658                                                                              & \multicolumn{1}{c|}{34}           & 513         \\ \hline
1991-2000                         & \multicolumn{1}{c|}{2,042}          & 6.91             & 7,88,617                                                                             & 45,757                                                                             & \multicolumn{1}{c|}{175}          & 1,867       \\ \hline
2001-2010                         & \multicolumn{1}{c|}{9,680}          & 32.75            & 33,79,239                                                                            & 1,85,869                                                                           & \multicolumn{1}{c|}{1,204}        & 8,476       \\ \hline
2011-2020                         & \multicolumn{1}{c|}{14,614}         & 49.45            & 48,98,374                                                                            & 3,25,447                                                                           & \multicolumn{1}{c|}{3,929}        & 10,685      \\ \hline
2021-2023                         & \multicolumn{1}{c|}{2,402}          & 8.13             & 3,79,250                                                                             & 24,875                                                                             & \multicolumn{1}{c|}{973}          & 1,429       \\ \hline
Total                             & \multicolumn{1}{c|}{29,556}         & 100              & 97,53,620                                                                            & 5,93,321                                                                           & \multicolumn{1}{c|}{6,327}        & 23,229      \\ \hline
\end{tabular}
\end{table}
\begin{longtable}{|c|l|c|c|c|c|c|}
\caption{List of papers with more than 90\% of self-citations.}
\label{tab:sc}\\
\hline
\textbf{S. No.} & \textbf{Paper Title}                                                                                                                                                                                       & \textbf{PubYear} & \textbf{\begin{tabular}[c]{@{}c@{}}Team\\ Size\end{tabular}} & \textbf{\begin{tabular}[c]{@{}c@{}}Total\\ Citations\end{tabular}} & \textbf{\begin{tabular}[c]{@{}c@{}}Self\\ Citations\end{tabular}} & \textbf{\begin{tabular}[c]{@{}c@{}}\% Self\\ Citations\end{tabular}} \\ \hline
\endhead
1               & \begin{tabular}[c]{@{}l@{}}Analysis, adaptive control\\ and synchronization of a\\ seven-term novel 3-D\\ chaotic system\end{tabular}                                                                      & 2013             & 2                                                            & 176                                                                & 173                                                               & 98.3                                                                 \\ \hline
2               & \begin{tabular}[c]{@{}l@{}}Global chaos synchronization\\ of a family of n-scroll hyperchaotic\\ chua circuits using backstepping\\ control with recursive feedback\end{tabular}                           & 2013             & 2                                                            & 148                                                                & 145                                                               & 98.0                                                                 \\ \hline
3               & \begin{tabular}[c]{@{}l@{}}Sliding controller design of\\ hybrid synchronization of\\ Four-Wing Chaotic systems\end{tabular}                                                                               & 2011             & 2                                                            & 182                                                                & 178                                                               & 97.8                                                                 \\ \hline
4               & \begin{tabular}[c]{@{}l@{}}Sliding mode control based\\ global chaos control of\\ Liu-Liu-Liu-Su chaotic\\ system\end{tabular}                                                                             & 2012             & 1                                                            & 165                                                                & 161                                                               & 97.6                                                                 \\ \hline
5               & \begin{tabular}[c]{@{}l@{}}Adaptive anti-synchronization\\ of Uncertain Tigan and Li\\ Systems\end{tabular}                                                                                                & 2012             & 2                                                            & 162                                                                & 158                                                               & 97.5                                                                 \\ \hline
6               & \begin{tabular}[c]{@{}l@{}}Active controller design for\\ generalized projective\\ synchronization of\\ four-scroll chaotic systems\end{tabular}                                                           & 2011             & 2                                                            & 164                                                                & 159                                                               & 97.0                                                                 \\ \hline
7               & \begin{tabular}[c]{@{}l@{}}A new eight-term 3-D\\ polynomial chaotic system\\ with three quadratic\\ nonlinearities\end{tabular}                                                                           & 2014             & 1                                                            & 181                                                                & 175                                                               & 96.7                                                                 \\ \hline
8               & \begin{tabular}[c]{@{}l@{}}Global chaos control of\\ hyperchaotic Liu system\\ via sliding control method\end{tabular}                                                                                     & 2012             & 1                                                            & 163                                                                & 157                                                               & 96.3                                                                 \\ \hline
9               & \begin{tabular}[c]{@{}l@{}}Anti-synchronization of LÃ¼\\ and Pan chaotic systems by\\ adaptive nonlinear control\end{tabular}                                                                              & 2011             & 2                                                            & 161                                                                & 155                                                               & 96.3                                                                 \\ \hline
10              & \begin{tabular}[c]{@{}l@{}}Global chaos synchronization\\ of hyperchaotic Pang and\\ hyperchaotic Wang systems\\ via adaptive control\end{tabular}                                                         & 2012             & 2                                                            & 152                                                                & 146                                                               & 96.1                                                                 \\ \hline
11              & \begin{tabular}[c]{@{}l@{}}A new six-term 3-D chaotic\\ system with an exponential\\ nonlinearity\end{tabular}                                                                                             & 2013             & 1                                                            & 197                                                                & 189                                                               & 95.9                                                                 \\ \hline
12              & \begin{tabular}[c]{@{}l@{}}Generalized Projective\\ Synchronization of Two-Scroll\\ Systems via Adaptive Control\end{tabular}                                                                              & 2012             & 2                                                            & 162                                                                & 155                                                               & 95.7                                                                 \\ \hline
13              & \begin{tabular}[c]{@{}l@{}}Anti-synchronization of\\ hyperchaotic lorenz and\\ hyperchaotic chen systems\\ by adaptive control\end{tabular}                                                                & 2011             & 2                                                            & 158                                                                & 151                                                               & 95.6                                                                 \\ \hline
14              & \begin{tabular}[c]{@{}l@{}}The generalized projective\\ synchronization of hyperchaotic\\ lorenz and hyperchaotic Qi\\ systems via active control\end{tabular}                                             & 2011             & 2                                                            & 165                                                                & 157                                                               & 95.2                                                                 \\ \hline
15              & \begin{tabular}[c]{@{}l@{}}Hybrid synchronization of\\ n-scroll chaotic chua circuits\\ using adaptive backstepping\\ control design with recursive\\ feedback\end{tabular}                                & 2013             & 2                                                            & 161                                                                & 152                                                               & 94.4                                                                 \\ \hline
16              & \begin{tabular}[c]{@{}l@{}}Analysis, properties and control\\ of an eight-term 3-D chaotic\\ system with an exponential\\ nonlinearity\end{tabular}                                                        & 2015             & 1                                                            & 154                                                                & 145                                                               & 94.2                                                                 \\ \hline
17              & \begin{tabular}[c]{@{}l@{}}Generalised projective synchronisation\\ of novel 3-D chaotic systems\\ with an exponential non-linearity\\ via active and adaptive control\end{tabular}                        & 2014             & 1                                                            & 166                                                                & 155                                                               & 93.4                                                                 \\ \hline
18              & \begin{tabular}[c]{@{}l@{}}Adaptive synchronization of \\ chemical chaotic reactors\end{tabular}                                                                                                           & 2015             & 1                                                            & 149                                                                & 136                                                               & 91.3                                                                 \\ \hline
19              & \begin{tabular}[c]{@{}l@{}}Global chaos synchronization\\ of WINDMI and Coullet chaotic\\ systems using adaptive\\ backstepping control design\end{tabular}                                                & 2014             & 2                                                            & 160                                                                & 146                                                               & 91.3                                                                 \\ \hline
20              & \begin{tabular}[c]{@{}l@{}}Analysis, control and synchronisation\\ of a six-term novel chaotic system\\ with three quadratic nonlinearities\end{tabular}                                                   & 2014             & 1                                                            & 192                                                                & 175                                                               & 91.1                                                                 \\ \hline
21              & \begin{tabular}[c]{@{}l@{}}Analysis and anti-Synchronization\\ of a novel chaotic system via active\\ and adaptive controllers\end{tabular}                                                                & 2013             & 1                                                            & 184                                                                & 167                                                               & 90.8                                                                 \\ \hline
22              & \begin{tabular}[c]{@{}l@{}}Analysis, adaptive control and\\ anti-synchronization of a six-term\\ novel jerk chaotic system with two\\ exponential nonlinearities and its\\ circuit simulation\end{tabular} & 2015             & 5                                                            & 150                                                                & 136                                                               & 90.7                                                                 \\ \hline
23              & \begin{tabular}[c]{@{}l@{}}Analysis and adaptive synchronization\\ of two novel chaotic systems with\\ hyperbolic sinusoidal and cosinusoidal\\ nonlinearity and unknown parameters\end{tabular}           & 2013             & 1                                                            & 191                                                                & 173                                                               & 90.6                                                                 \\ \hline
\end{longtable}

\subsection{Team size vs. self-citations}

\begin{figure}[!h]
    \centering
\includegraphics[width=0.95\linewidth]{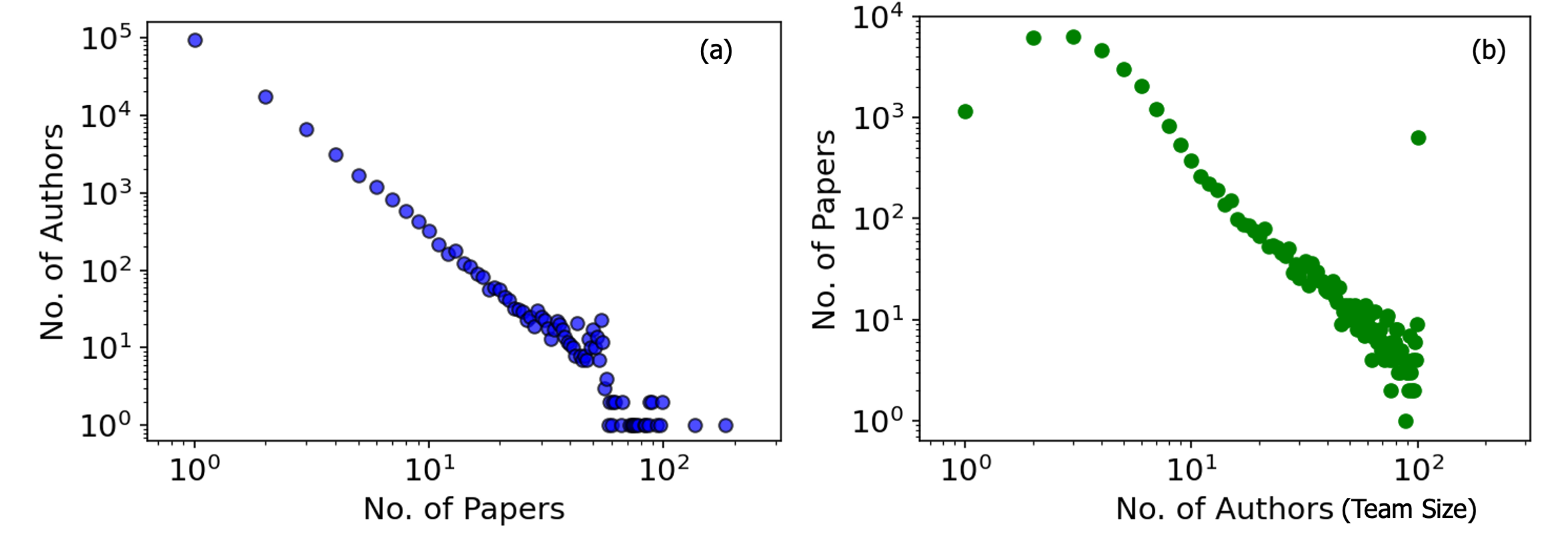} 
\caption{(a) Number of papers vs. count of authors. (b) Team size vs. number of papers.}
\label{fig:authPapers}
\end{figure}

According to \cite{wuchty2007increasing}, larger teams tend to dominate in producing highly cited research due to their ability to combine various expertise, tackle complex problems, and leverage collaborative networks. In contrast, smaller teams are more likely to focus on niche and innovative topics, which may gain recognition more gradually. This highlights the role of team size in shaping research impact and citation patterns. Figure~\ref{fig:authPapers}(a) illustrates the authors with multiple highly cited papers. 73.47\% of the authors have a single paper appearing in the top 1\% highly cited papers. 13.55\% authors have two papers in the highly cited list followed by 5.2\% having 3 papers, etc. \color{black} Figure~\ref{fig:authPapers}(b) represents the distribution of teams appearing in the paper and the number of publications. A negative correlation exists between the number of papers and the number of authors, suggesting that smaller teams (fewer authors) are more prolific in producing papers. In contrast, larger teams contribute fewer papers, likely due to the increased complexity and coordination involved in collaborative research. As team size grows, the number of papers decreases significantly, with some exceptions for very large teams.

\begin{figure}[!h]
    \centering
\includegraphics[width=0.75\linewidth]{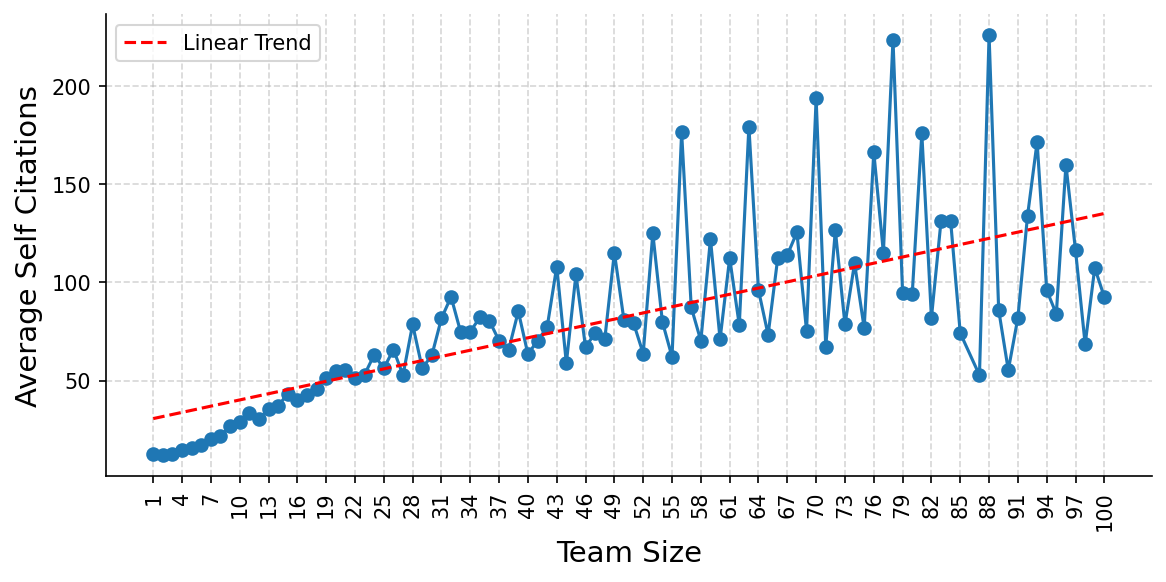} 
\caption{Team size vs average self-citations.}
\label{fig:Team_Avg_SC}
\end{figure}

 In addition, Figure~\ref{fig:Team_Avg_SC} represents the average self-citations based on team size. The average self-citations generally increase as the team size grows, particularly in smaller teams. There are notable peaks and troughs, indicating variability in self-citation practices as team sizes change. Teams with fewer members (1–25) exhibit a more consistent, gradual increase in average self-citations, indicating relatively steady behavior. The red dashed line illustrates the overall increasing trend, suggesting a positive correlation (1.05) between team size and average self-citations, although the data show variability.

\subsection{Collaboration pattern vs. self-citations}
\begin{table}[]
\caption{Distribution of number of papers, corresponding total and self-citations.}
\label{tab:inter-intra}
\begin{tabular}{|l|ll|ll|l|}
\hline
\multirow{2}{*}{\textbf{Category}} & \multicolumn{2}{c|}{\textbf{Domestic}}               & \multicolumn{2}{c|}{\textbf{International}}          & \multicolumn{1}{c|}{\multirow{2}{*}{\textbf{Total}}} \\ \cline{2-5}
                                   & \multicolumn{1}{l|}{\textbf{Count}} & \textbf{In \%} & \multicolumn{1}{l|}{\textbf{Count}} & \textbf{In \%} & \multicolumn{1}{c|}{}                                \\ \hline
No. of papers                      & \multicolumn{1}{l|}{16,823}         & 56.9\%         & \multicolumn{1}{l|}{12,733}         & 43.1\%         & 29,556                                               \\ \hline
No. of total citations             & \multicolumn{1}{l|}{48,43,057}      & 49.7           & \multicolumn{1}{l|}{49,10,563}      & 50.3\%         & 97,53,620                                            \\ \hline
No. of self citations              & \multicolumn{1}{l|}{1,95,752}       & 33\%           & \multicolumn{1}{l|}{3,97,569}       & 67\%           & 5,93,321                                             \\ \hline
\end{tabular}
\end{table}
An influential and highly cited paper is often the result of interdisciplinary teams and collaborative efforts. According to \citet{uzzi2013atypical}, interdisciplinary collaboration and diversity in research teams are significant factors contributing to innovative and impactful research, as they bring together varied perspectives and expertise, which increase the likelihood of producing groundbreaking work.
The table~\ref{tab:inter-intra} provides an overview of the distribution of highly cited papers and self-citations between domestic and international collaborations. Out of the total 29,556 highly cited papers, domestic collaborations contribute the majority, accounting for 56.9\%, while international collaborations contribute 43.1\%. The total citation distribution indicates that 49.7\% of total citations belong to domestic papers, while 50.3\% of the total citations belong to international papers. However, when examining self-citations, international collaborations dominate with 67\% self-citations, compared to 33\% self-citations from domestic collaborations. This indicates that while domestic collaborations produce a higher number of highly cited papers, international collaborations are more likely to generate self-citations. This could reflect the broader scope, larger teams, and multidisciplinary nature of international projects, leading to higher interconnectedness and frequent self-referencing. In contrast, domestic collaborations, often involving smaller teams, focus on national-level research with relatively fewer self-citations.

\begin{figure}[!h]
    \centering
\includegraphics[width=0.49\linewidth]{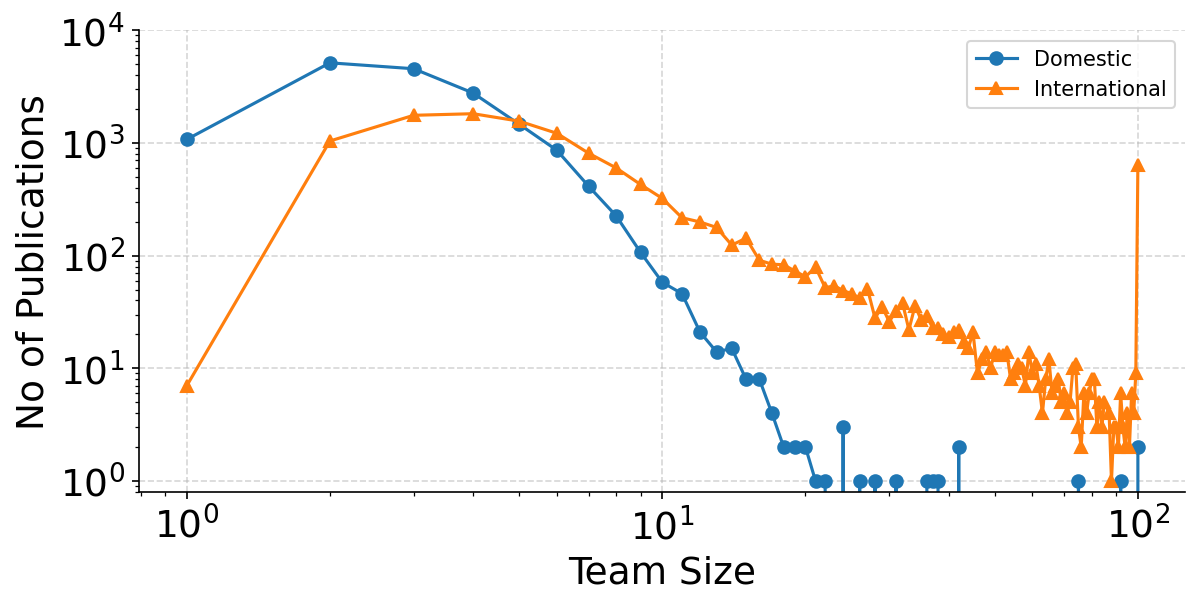} 
\includegraphics[width=0.49\linewidth]{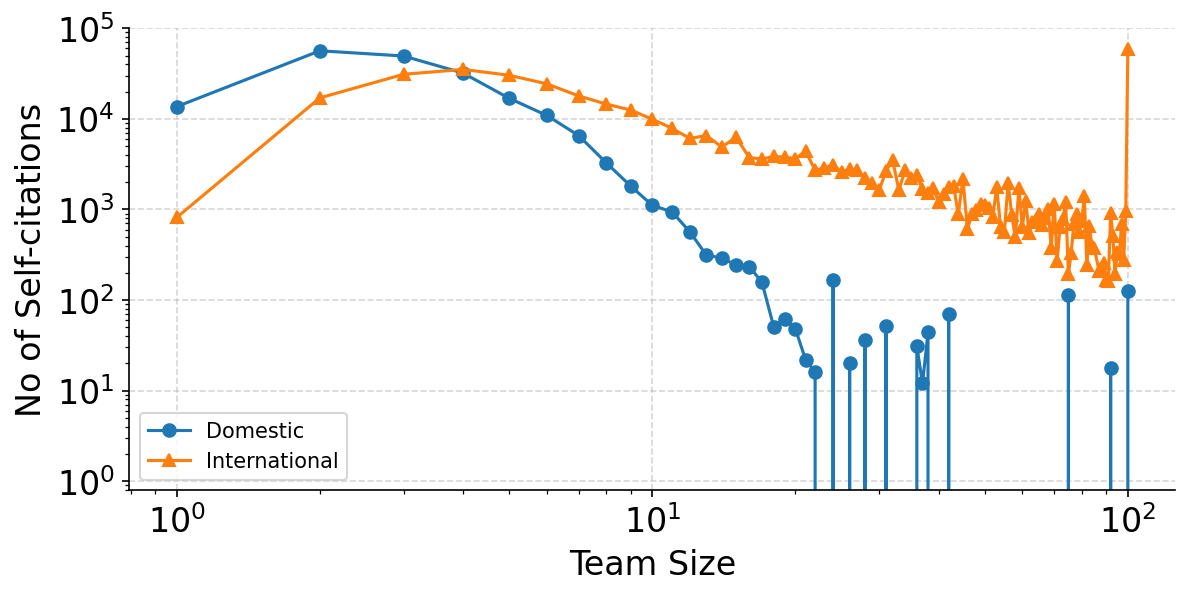} 
\caption{Domestic vs. international collaboration. (left) Team size vs. number of publications. (b) Team size vs. self-citations.}
\label{fig:teamCollab}
\end{figure}

Figure~\ref{fig:teamCollab} represents different aspects of the relationship between team size and the nature of collaborations (domestic vs. international).
The figure on the left represents the number of publications versus team size where publications for domestic collaborations peak at smaller team sizes and gradually decline as team size increases, showing minimal activity for larger teams. International collaborations demonstrate a consistent trend, maintaining a higher level of publications for medium-to-large team sizes compared to domestic collaborations. The spike in international collaborations for the largest team size is likely driven by highly collaborative or global-scale projects.
Similarly, the right figure represents the number of self-citations vs. team size where self-citations for domestic collaborations are higher for smaller team sizes and decline sharply as the team size increases. International collaborations exhibit a steadier decline in self-citations, with smaller team sizes showing relatively lower self-citations compared to domestic ones.
There is a noticeable peak for international collaborations at the larger team size (likely an outlier). In general, domestic collaborations dominate smaller team sizes in terms of both self-citations and publications, whereas international collaborations gain prominence as team sizes increase. In addition, the USA is the topmost collaborator followed by the UK and Germany.


\section{Conclusion}

In conclusion, citations and self-citations play a crucial role in academic research by measuring impact, ensuring research continuity, and fostering collaboration. The study on highly cited Indian research papers highlights that domestic collaborations contribute a greater number of highly cited papers, while international collaborations generate more self-citations, likely due to broader networks and multidisciplinary projects. Team size also influences citation patterns, with smaller teams producing more papers but relying more on self-citations, whereas larger teams are linked to international collaborations and higher citation visibility. Striking a balance between self-referencing and engaging with the broader academic community is essential to maintain credibility and ensure meaningful research impact.

\subsection{Limitations}

This study has several limitations that should be considered when interpreting the findings. First, citation practices vary across disciplines, making direct comparisons challenging, especially in fields where self-citation rates are naturally higher. Second, limitation lies in the classification of team sizes, as it does not account for variations in individual author contributions, which can influence citation impact. Third, the study differentiates between domestic and international collaborations but does not fully capture the complexity of multi-country partnerships that may affect citation trends. Lastly certain fields, such as biomedical research, have higher citation frequencies than others, making direct comparisons across disciplines challenging.

\section*{Data availability}
The datasets used in the study will be available from the corresponding author on request.

\section*{Acknowledgment}
Authors gratefully acknowledges the Research and Development Cell, BML Munjal University for
their financial support through the seed grant (No: BMU/RDC/SG/2024-06), which made this research possible.

 \section*{Conflict of interest}
 The authors declare no conflict of interest.
\bibliographystyle{cas-model2-names}
\bibliography{cas-refs}

\end{document}